\documentclass[smallabstract,smallcaptions]{dccpaper}
\usepackage{graphicx}
\usepackage{epsfig}
\usepackage{amsmath}
\usepackage{amssymb}
\usepackage{color}
\usepackage{url}

\newlength{\figurewidth}
\newlength{\smallfigurewidth}

\begin{document}

\title
{\large
\textbf{CNN-based driving of block partitioning for intra slices encoding}
}

\author{%
Franck Galpin, Fabien Racap\'{e}, Sunil Jaiswal,\\Philippe Bordes, Fabrice Le L\'{e}annec and Edouard Fran\c{c}ois\\[0.5em]
{\small\begin{minipage}{\linewidth}\begin{center}
\begin{tabular}{ccc}
Technicolor \\
975 Avenue des Champs Blancs \\
35576, Cesson-S\'{e}vign\'{e}\\
\url{firstname.lastname@technicolor.com} 
\end{tabular}
\end{center}\end{minipage}}
}

\maketitle
\thispagestyle{empty}

\begin{abstract}
This paper provides a technical overview of a deep-learning-based encoder method aiming at optimizing next generation hybrid video encoders for driving the block partitioning in intra slices. An encoding approach based on Convolutional Neural Networks is explored to partly substitute classical heuristics-based encoder speed-ups by a systematic and automatic process. The solution allows controlling the trade-off between complexity and coding gains, in intra slices, with one single parameter. This algorithm was proposed at the Call for Proposals of the Joint Video Exploration Team (JVET) on video compression with capability beyond HEVC. In All Intra configuration, for a given allowed topology of splits, a speed-up of $\times 2$ is obtained without BD-rate loss, or a speed-up above $\times 4$ with a loss below 1\% in BD-rate.
\end{abstract}

\section{Introduction}

Neural Networks (NN) have recently shown promising results, when designed to partly or fully replace image codecs. As the introduction of NN-based techniques currently represent a huge challenge regarding computer complexity issues, this paper focuses on the optimization of the encoder side only. In particular, this work tackles the partitioning of images into blocks of different sizes for optimizing their compression. 

Considering the evolution of the last video compression standards, one can observe that an increased coding efficiency has been obtained by extending the picture partitioning options. In HEVC/H.265 \cite{HEVC}, the basic partitioning of a non-overlapping $64\times64$ Coding Tree Units (CTU) is based on a recursive quad-tree (QT) decomposition into smaller CUs. In the Joint Exploration Model (JEM) developed by the Joint Video Exploration Team (JVET), the concept was extended by the introduction of Binary Tree splitting (BT) that allows splitting a Coding Unit (CU) into two symmetric rectangular sub-CUs. Although the standards only specify the syntax that signals the partitioning of these CUs, encoders require an efficient way to choose and optimize the sizes of blocks over the images, depending on its ability to compress the different regions. This process is included into the so-called Rate-Distortion Optimization (RDO) that aims at finding the best compromise between a target quality of a reconstructed region and the required bits to be transmitted. 


Classical encoder speed-ups, usually based on the analysis of the behavior of a full RDO, are used to avoid a fully exhaustive search over all possible combinations of CU partition, coding mode, prediction mode and transform type. These fast methods are applied at the cost of approximations.
Deep-learning based algorithms offer a complementary way to address the combinatorics issue, with the possibility to develop automatically trained engines. 

An approach relying on Convolutional Neural Networks (CNN) was proposed by Liu et al. in \cite{liu_cnn_2016} to drive the encoder for deciding on whether to split a given block, within the framework of HEVC. All blocks from $8\times 8$ to $32\times 32$ are first sampled to an $8\times 8$ matrix by averaging, providing an $8\times 8$ matrix as input to the CNN, which outputs a duplet of values indicating to split a current block into four smaller blocks or not (quadtree). A deeper approach was proposed in \cite{DBLP:journals/corr/abs-1710-01218} to optimize the partitioning of HEVC CUs using a partially convolutional approach on the Luma component only. More recently, classical machine learning based approaches \cite{Sun2018AFI} also reach competitive results for intra splits prediction on HEVC. Results of these approaches are further discussed in section \ref{sec:results}.

This paper describes a deep-learning-based approach implemented within the proposed encoder of \cite{J0022}, submitted as a response to the call for proposal~\cite{h1002} on video compression with capability beyond HEVC \cite{HEVC}, organized by the Joint Video Exploration Team (JVET). This approach aims at driving the encoder by estimating probabilities of blocks or Coding Units (CU) splitting in intra slices using the content of these blocks. The approach is primarily based on a texture analysis of the original blocks, and partly replaces the costly RDO potentially involved for testing all potential partitioning configurations. Compared to other methods, the proposed approach enables to consider multiple types of splits: QT, BT and introduced Asymmetric Binary Tree (ABT). It also considers the introduced dual tree, which separate tree structures for the luma and chroma components. 


The remainder of this paper is organized as follows. The overview of the proposed method is presented in section \ref{sec:overview}. Section \ref{sec:cnn} details the CNN architecture and training. Then, the processing of split probabilities is explained in section \ref{sec:split_selection}. Finally, the implementation and results of the proposed method are discussed in section \ref{sec:results}. 

%

\section{Algorithm overview}\label{sec:overview}
A Convolutional Neural Network (CNN) based analysis is proposed to speed-up the intra slices encoding process. The method is based on texture analysis of original $64\times 64$ root blocks, luma or chroma, for predicting the most probable splits inside each potential sub-block. The method aims at enabling the encoder to leverage the gains of BT and ABT partitioning, while reducing the combinations for the RDO checks. 
The solution described in \cite{J0022} considers CTUs of size $256\times 256$. In intra slices, the encoder always applies a quadtree split to the CTU as a first splitting depth, leading to four $128\times 128$ blocks in luma and $64\times 64$ blocks in chroma. In the luma case, the second splitting depth only allows quadtree split, evaluated through a classical RDO process, leading to one $128\times 128$ or four $64\times 64$ blocks. Root blocks for CNN based analysis then correspond to the third level of depth in luma, respectively the second level in chroma. $64\times 64$ blocks are considered in both cases.  
In fact, $65\times 65$ samples patches, luma or chroma, are considered as input to the CNN stage. They consist in the original texture of root blocks plus their causal neighboring top and left boundary samples, as they participate to split decisions in classical RDO. 

\begin{figure}[htp]
\begin{center}
\epsfig{width=6in,file=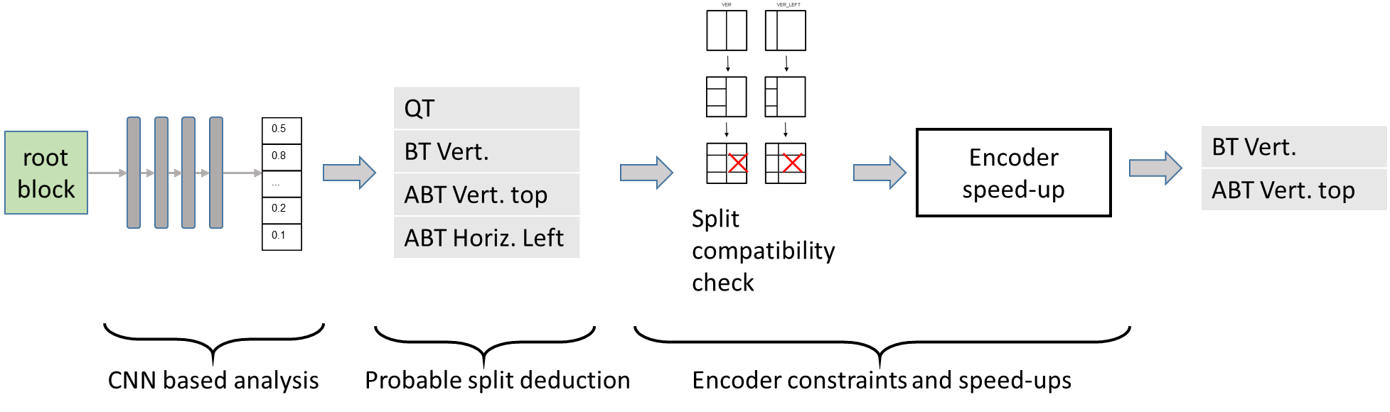} 
\end{center}
\vspace{-0.5in}
\caption{\label{fig:overview}%
Overview of the split prediction process.}
\vspace{-0.25in}
\end{figure}

Fig. \ref{fig:overview} depicts the overall process of the split prediction module. This module precedes the usual RDO process, and pre-selects the split configurations to be tested by the RDO. It is composed of the following 3 steps:
\begin{enumerate}
\item \textbf{CNN-based analysis --} In the first step, each input $65\times 65$ patch is analyzed by a CNN-based texture analyzer. The output consists of a vector of probabilities associated to each elementary boundary that separate elementary sub-blocks. Fig.~\ref{fig:vector} illustrates the mapping between elementary boundary locations and the vector of probabilities. The size of elementary blocks being $4\times 4$, the vector contains $n=480$ probability values. The CNN is described in section \ref{sec:cnn}.

\begin{figure}[htp]
\begin{center}
\vspace{-0.2in}
\epsfig{width=3in,file=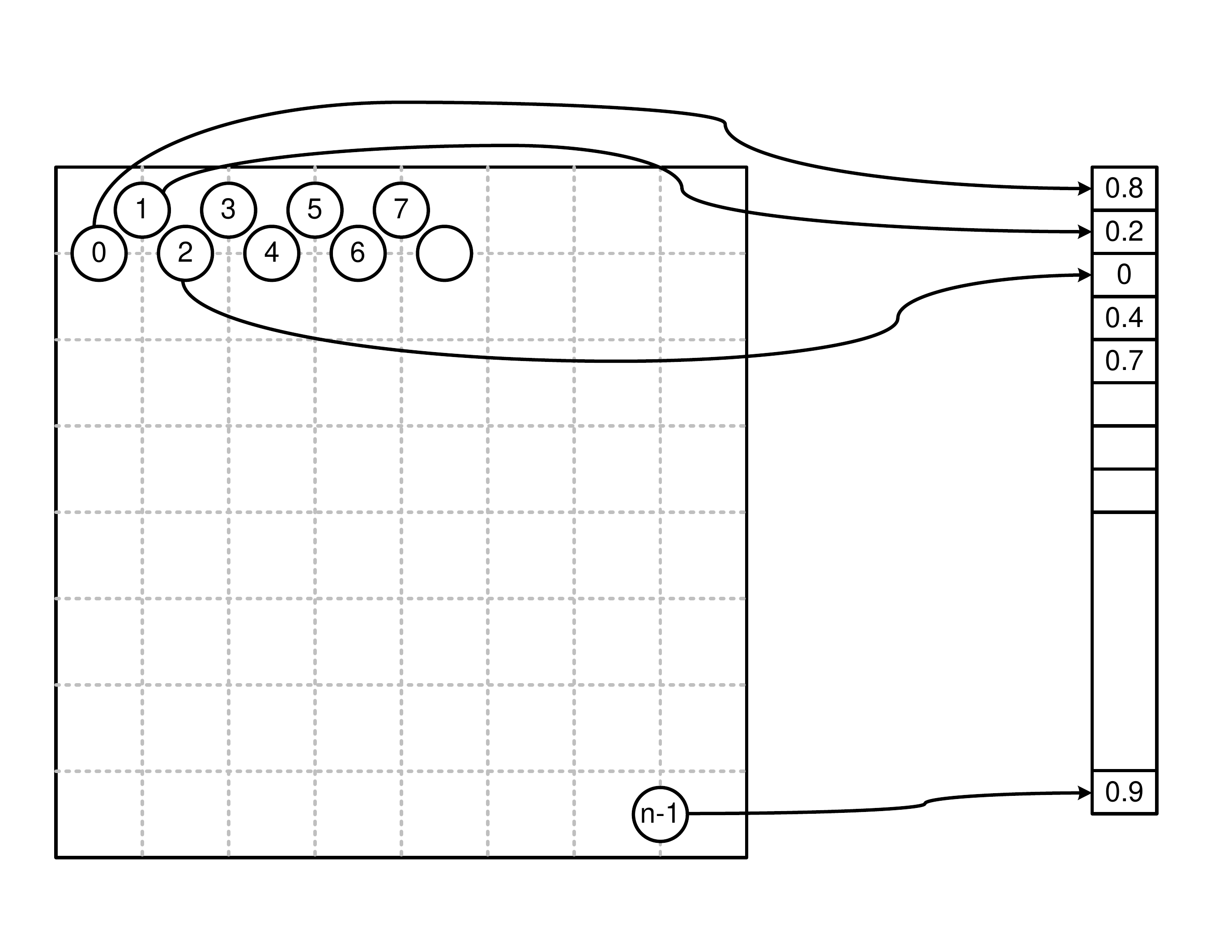}
\end{center}
\vspace{-0.6in}
\caption{\label{fig:vector}%
Mapping of the boundary locations onto a vector $n\times 1$.}
\vspace{-0.25in}
\end{figure}

\item \textbf{Probable split selection -- }  The second step takes as input the probability of each elementary boundary and outputs a first set of splitting modes among all possible options, which are: no split, QT, BT (vertical, horizontal), ABT (top, bottom, left, right). This step is further detailed in section \ref{sec:split_selection}.

\item 	\textbf{Encoder constraints and speed-ups --}   The third step selects the final set of splitting modes to be checked by classical RDO, depending on the first set provided by step 2, the contextual split constraints and the encoder speed-ups described in \cite{J0022}. This step is further detailed in section \ref{sec:enc_constraints}.
\end{enumerate}

\section{CNN-based analysis}\label{sec:cnn}

\subsection{CNN architecture}

Two CNNs are used, one dedicated to the luma component and one for the chroma components. They are trained and inferred independently, since luma and chroma blocks may have separate partitions in intra slices.

The CNN is loosely based on a small ResNet as described in \cite{he2016deep}, composed of a set of convolutional layers with several skip connections. The main differences with a small ResNet (i.e. 18-layer version with 2 building blocks) are:

- the adaptation of the number of filters (e.g. up to 48 filters instead of 512) and layers (13 layers instead of 18),

- the absence of batch norm, average pooling and double fully connected layers,

- the pooling at the end of each scaling block,

- the absence of stride during convolution.

- the model contains $\sim 225k$ parameters (compared to $\sim 10M$ for ResNet-18).

\begin{figure}[htp]
\begin{center}
\begin{scriptsize}
\def\svgwidth{\columnwidth}
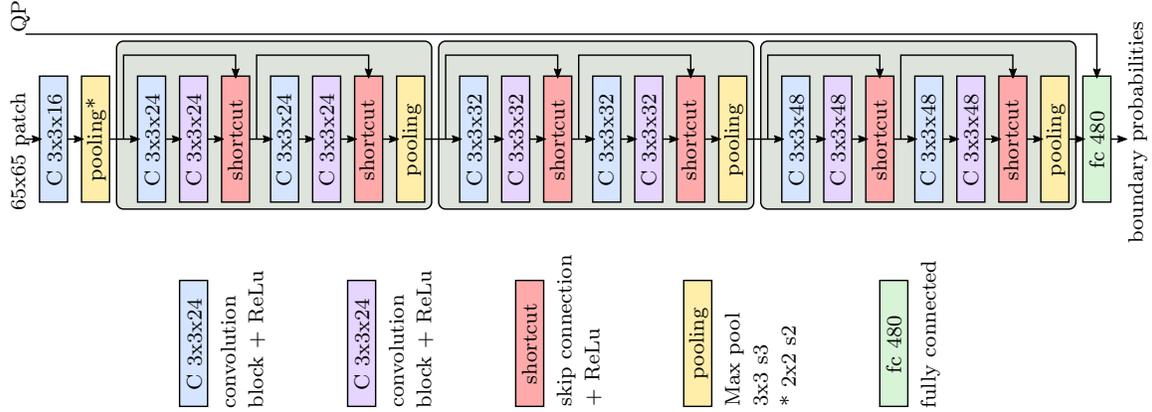
\end{scriptsize}
\end{center}
\vspace{-0.3in}
\caption{\label{fig:network}%
Overview of the CNN layout for boundary prediction.}
\vspace{-0.25in}
\end{figure}

The luma CNN takes as inputs a block of luma samples of size $65\times 65$, as well as a normalized luma Quantization step (Qstep) at the very end during the fully connected layer. The chroma CNN considers a chroma block $65\times 65\times 2$ and a normalized chroma Qstep at the very end during the fully connected layer.

It outputs a vector describing the probability of the 480 elementary boundary locations in the input root block as described in Fig. \ref{fig:vector}. 

\subsection{CNN training}

The training set was created by encoding a set of images in intra mode with the codec proposed in \cite{J0022}, using a classical RDO. A training set containing a large number of source images, excluding the JVET test sequences, was created. It is composed mainly of images from the Div2K dataset and some other internal database. It contains more than 10M patches, encoded with QPs ranging from 19 to 41. 

\begin{figure}[htp]
\begin{center}
\epsfig{width=2.5in,file=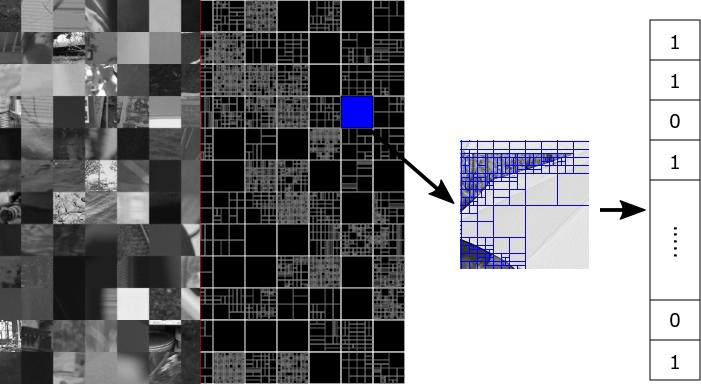} 
\end{center}
\vspace{-0.35in}
\caption{\label{fig:dataset}
Dataset creation for CNN training.}
\vspace{-0.25in}
\end{figure}

The ground truth corresponds to the split choices made by the RDO. During this encoding, several speed-ups, based on heuristics, were disabled, leading to a higher number of potential partitions compared to the actual codec version used in \cite{J0022}. For each block $64\times 64$, a patch of size $65\times 65$ containing the $64\times 64$ block + 1 pixel-wide causal border was extracted, as depicted in Fig.~\ref{fig:dataset}.
The Qstep used for encoding was also extracted for each block and normalized to consider a value ranging between 0 and 1 for the whole Qstep range.
Finally, the ground truth of each elementary boundary was extracted from the decoded bitstream and mapped onto a vector of size 480 using the inverse process as shown in Fig.\ref{fig:dataset}.

The loss function corresponds to a $L_2$ norm between the ground truth vector $V_1$, which contains values 1 or 0 depending on whether a boundary exists or not, and the output of the network $V_i$, plus a regularization term corresponding to the $L_2$ norm of the weights of the convolution $c_k$:
\begin{equation}
\mathcal{L} = \sum_{i}{\|{V_1}-V_i\|}+\lambda\sum_k{\|c_k\|}
\end{equation}

The CNN is classically trained using the deep learning framework Tensorflow \cite{abadi2016tensorflow}. Specifically, the Adam minimizer was used with a learning rate of $10^{-3}$, a regularization weight of $10^-5$ and a batch size of 256. The model was trained on 4 epochs in a few hours on a single recent GPU (GTX-1070).

\section{Probable split selection}

\subsection{Split selection}\label{sec:split_selection}
\begin{figure}[htp]
\begin{center}
\epsfig{width=5in,file=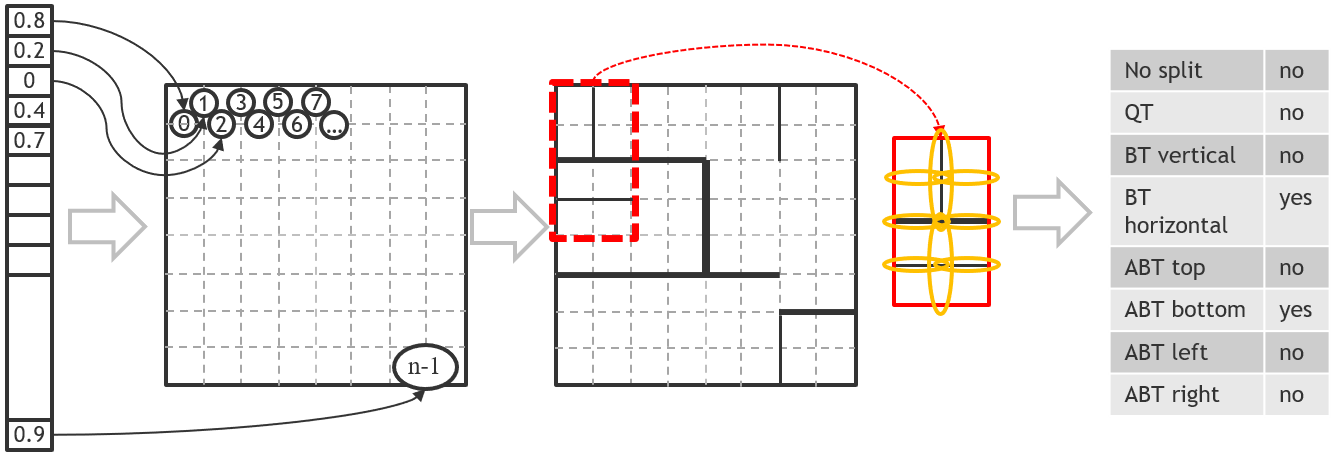}
\end{center} 
\vspace{-0.25in}
\hskip 0.5em a \hskip 7em  b \hskip 10em c \hskip 13em d
\vspace{-0.25in}
\caption{\label{fig:vector_to_decisions}%
From boundaries to split selection.}
\vspace{-0.25in}
\end{figure}

From the output vector of the CNN, an image of boundaries is constructed, as shown in Fig.~\ref{fig:vector_to_decisions} (a). For each sub-block potentially explored by the RDO, the relevant boundaries are extracted, represented as segments in Fig.~\ref{fig:vector_to_decisions} (b, c). 
At each location of possible splits, the average value of elementary boundaries corresponding to each half-split boundary is computed, as depicted in yellow in Fig.~\ref{fig:vector_to_decisions} (c). These values are then compared to pre-determined thresholds for each split choice. A decision to explore or not each choice is finally made. Fig.~\ref{fig:vector_to_decisions} (d) provides an example where, at this step, i.e. before performing the third step, QT, BT horizontal, and ABT bottom will be considered by the classical RDO.
The thresholds depend on the depth and a speed-control parameter described in section \ref{sec:tradeoff}.

%

The decision logic on each split depends on the type of partitioning choice. For example, when only QT is considered, either no split is explored if all corresponding boundary probabilities are lower than a pre-determined threshold, or Quadtree split is explored if at least one vertical and one horizontal half boundary probabilities are higher than a pre-determined threshold.
The remaining combinations, for RDO exploration, directly depend on the above thresholds. Note that the thresholds can be different for the luma tree and the chroma tree, enabling the encoder to make different trade-offs for each tree.




\subsection{Encoder constraints and speed-ups}\label{sec:enc_constraints}

This last step selects the final set of splitting modes to be checked by classical RDO.
Some splitting modes may be removed to be consistent with encoder heuristics, for example when they are not compatible with the current tree structure, e.g. quadtree emulation with an horizontal and a vertical split, when a splitting mode has already been explored in a previous configuration, or when no remaining splitting mode is available, the RDO falls back to the no-split mode exploration.

\section{Speed-control parameter and results}\label{sec:results}

This section details the combinations reduction of the RDO and describes how it is driven by a single parameter. It also provides results in terms of BD-rate gains versus encoder complexity.

\subsection{Raw trade-off speed/performance}

\begin{figure}[htp]
\begin{center}
\epsfig{width=5in,file=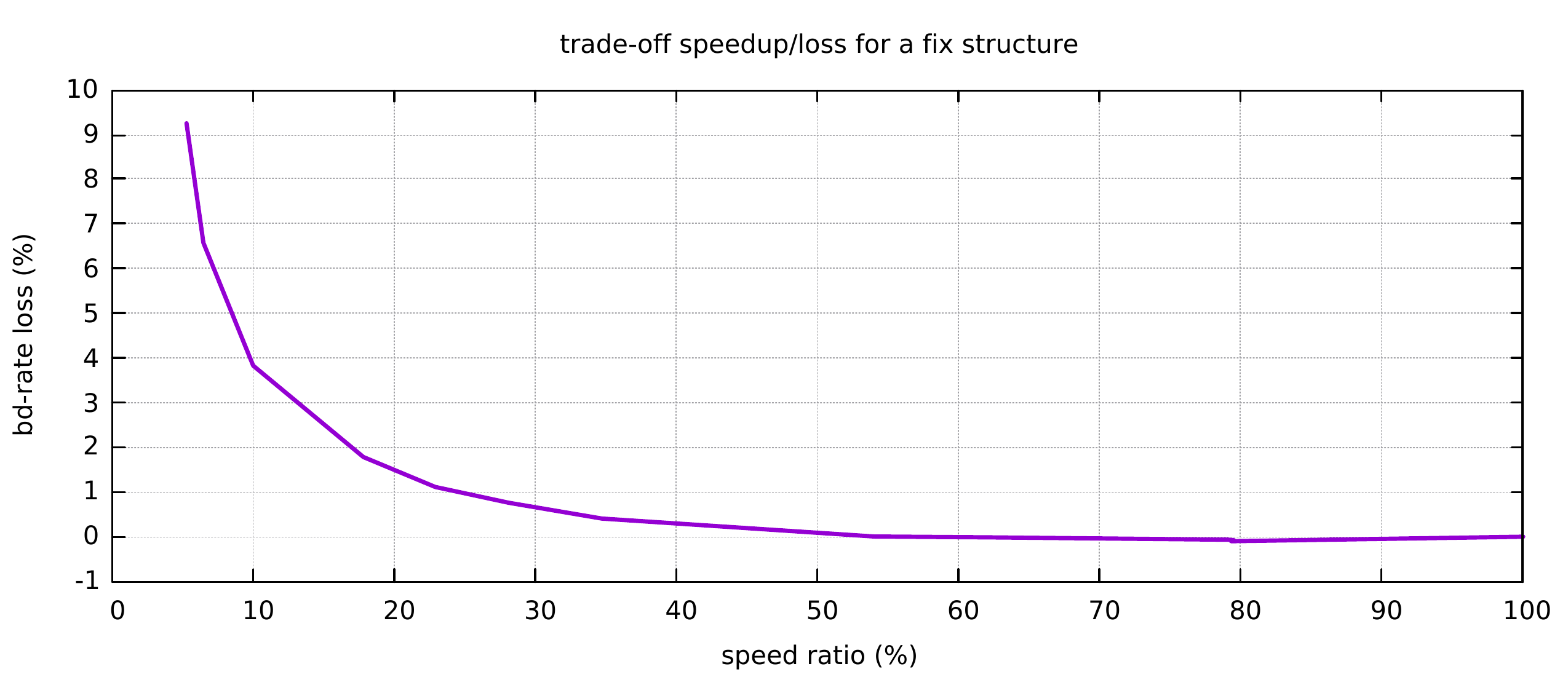} 
\end{center}
\vspace*{-1cm}\caption{BD-rate loss as a function of the encoding speed-up for a fix configuration.}\label{fig:tradeoff}%
\vspace{-0.25in}
\end{figure}

Fig~\ref{fig:tradeoff} shows an example of trade-off obtained using the proposed method. In this example, the codec tree structure capabilities have been fixed to the following:

- the maximum QT depth is 3 ($32 \times 32$),

- the ABT/BT depth at depth 2 ($64 \times 64$) is 2,

- the ABT/BT depth at depth 3 ($32 \times 32$) is 8.


The anchor to compare speed and performance is the codec without activating the CNN but activating standard codec heuristics for speed-up. The decision threshold is then changed and for each point it gives a BD-rate loss and a speed-up compared to anchor. One can see that a $\times 2$ speed-up is possible without BD-rate loss. For a loss below 1\% in BD-rate, the speed-up is above $\times 4$, while for a $\times 10$ speed-up, the BD-rate loss is about $3.8\%$.
\begin{table}[htp]
\begin{center}
\caption{\label{tab:results}%
Results on HM. 
}
{
\begin{footnotesize}
\begin{tabular}{|c|c|c|c|c|}
\hline
Method & algo.  & $\Delta$ bdrate  & $\Delta T$ \\
\hline
Liu 2016\cite{liu_cnn_2016} &   CNN         & $\sim 1.1\%$ & $27.9\%$ \\

Lu 2017 \cite{lu_dcc_2017}  & heuristics  & $\sim 2\%$   & $57.2\%$ \\

Xu 2017\cite{DBLP:journals/corr/abs-1710-01218} & CNN  & $\sim 1.4\%$ & $\sim 66\%$\\

Sun 2018\cite{Sun2018AFI}  & Canny+SVM &  $\sim 1.1\%$  & $48.6\%$ \\

Proposed  & CNN &  $0.005\%$  & $54\%$ \\
\hline
\end{tabular}
\end{footnotesize}
}
\end{center}
\end{table}
The table \ref{tab:results} summarizes some of related state-of-the-art results, although implemented for HEVC where the combinations are quite reduced compared to next generation codec. The comparison is then difficult with the presented approach. Moreover, the results presented here are taking into account all processing including the CNN run-time (no GPU used, see section \ref{sec:implementation}), whereas it is not always specified in previous state-of-the-art methods. 

\subsection{Encoder constraints and speed-ups}

Table~\ref{tab:speedup} depicts the estimate of the combinatorics for coding a single luma block of size $32\times 32$, using the full combinations for each tree structure using three software versions of the codec. 

\begin{table}[htp]
\begin{center}
\caption{\label{tab:speedup}%
\vspace*{0.25cm}Number of luma sub-blocks checked at the encoder side.}
{
\begin{footnotesize}
\begin{tabular}{|c|c|c|c|}
\hline
Tree structure & (almost) no heuristics & with heuristics & with CNN \\
\hline
QT & 8340 & 8181 (-2\%)& not tested\\
BT & 186932 ($\times 22$/QT) & 58388 (-69\%) & 49769 (-73\%) \\
BT+ABT & 577689 ($\times 3$/QTBT) & 416281 (-28\%) & 182156 (-68\%) \\
\hline
\end{tabular}
\end{footnotesize}
}
\end{center}
\end{table}
\vspace{-0.25in}

The first version corresponds to the encoding with almost no heuristics. The second version activates  classical encoding heuristics (as described in \cite{J0022}). The third version, in addition, implements the CNN-based algorithm. The benefit in terms of reduction of number of RD checks, resulting from the CNN approach, is observed in Table~\ref{tab:speedup}, e.g. $-68\%$ instead of $-28\%$ with heuristics for BT+ABT.

\subsection{Tree exploration trade-offs}\label{sec:tradeoff}

A single control parameter called speed-control has been added to the encoder. This parameter is a floating value. It drives the structure of the tree used for the luma and chroma: (maximum and minimum QT level and BT /ABT levels, as well as the thresholds used to prune the tree before the RDO.
Table~\ref{tab:treestructures} reports the link between the speed-control parameter and the maximum QT depth and the maximum number of BT/ABT splits. 

\begin{table}[htp]
\begin{center}
\caption{\label{tab:treestructures}%
Tree structures explored depending on the speed-control parameter.
}
\begin{footnotesize}
{
\begin{tabular}{|c|c|c|c|c|}
\hline
\begin{tabular}{c} speed-control \\ parameter \end{tabular}	& \begin{tabular}{c} Max \\ QT depth \end{tabular} & \begin{tabular}{c} ABT/BT \\ depth[2]($64\times 64$)  \end{tabular} & \begin{tabular}{c} ABT/BT  
\\depth[3]($32\times 32$)  \end{tabular} & \begin{tabular}{c} ABT/BT \\depth[4]($16\times 16$)  \end{tabular}\\
\hline
$\left[ 0.65,1.04\right[$ &	2 ($64\times 64$) &	8 & 0 &	0 \\
$\left[ 1.04,1.7\right[$ &	3 ($32\times 32$)&  2 &	8 &	0 \\
$\left[ 1.7,3.2\right[$ &	4 ($16\times 16$) &	0 &	4&	8 \\
$\left[ 3.2,-\right[$ &	3 ($32\times 32$) &	0 &	8 &	0 \\
\hline
\end{tabular}
}
\end{footnotesize}

\end{center}
\end{table}

In the following, some results are reported using the CfP test set, in AI mode, using the codec of \cite{J0022} including the ABT mode. Encoding and performance measurements were made on the first frame of each sequence.

\begin{figure}[htp]
\begin{minipage}[b]{0.52\textwidth}
    \includegraphics[width=\textwidth]{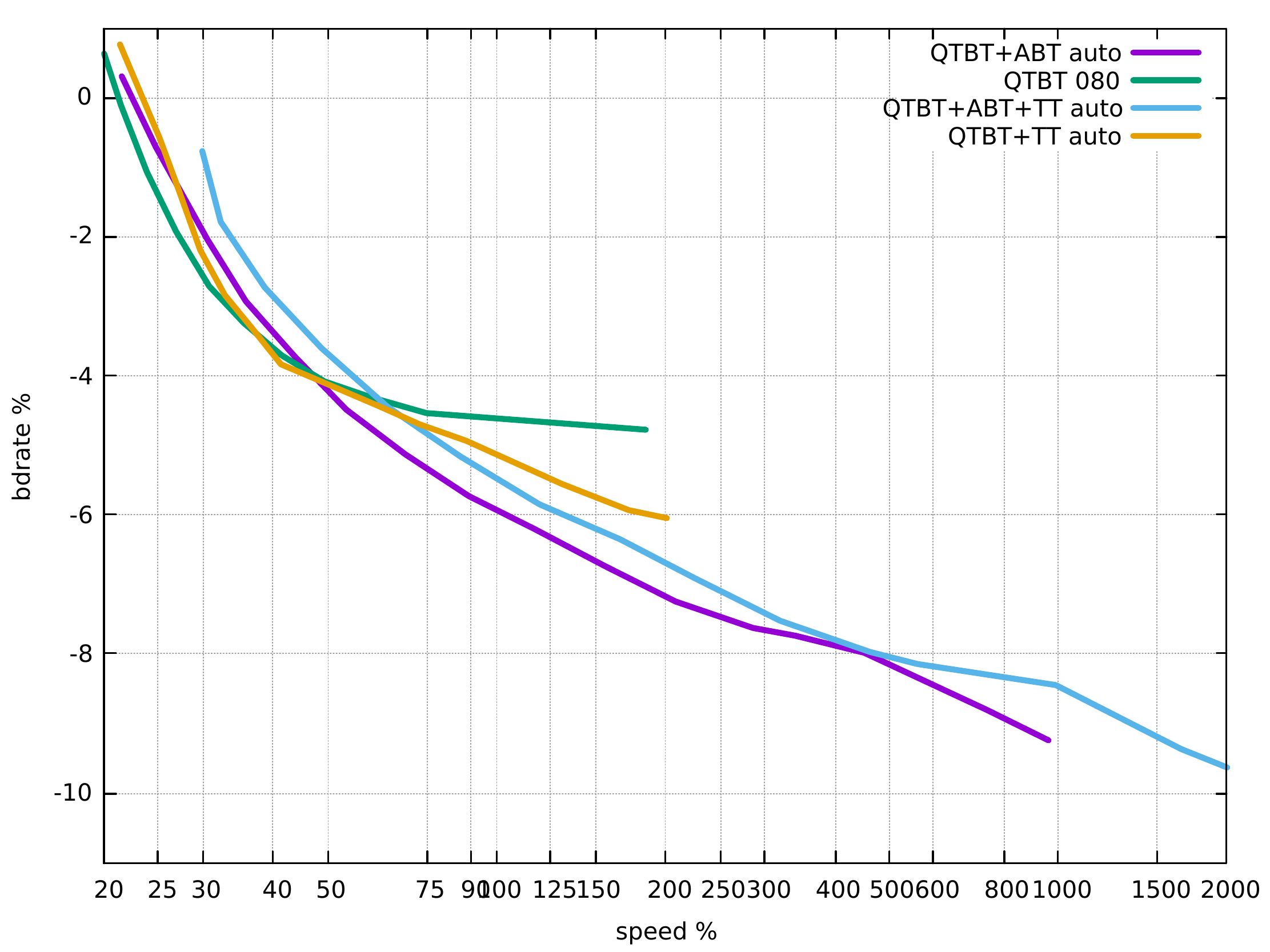}
  \end{minipage}
  \begin{minipage}[b]{0.52\textwidth}
    \includegraphics[width=\textwidth]{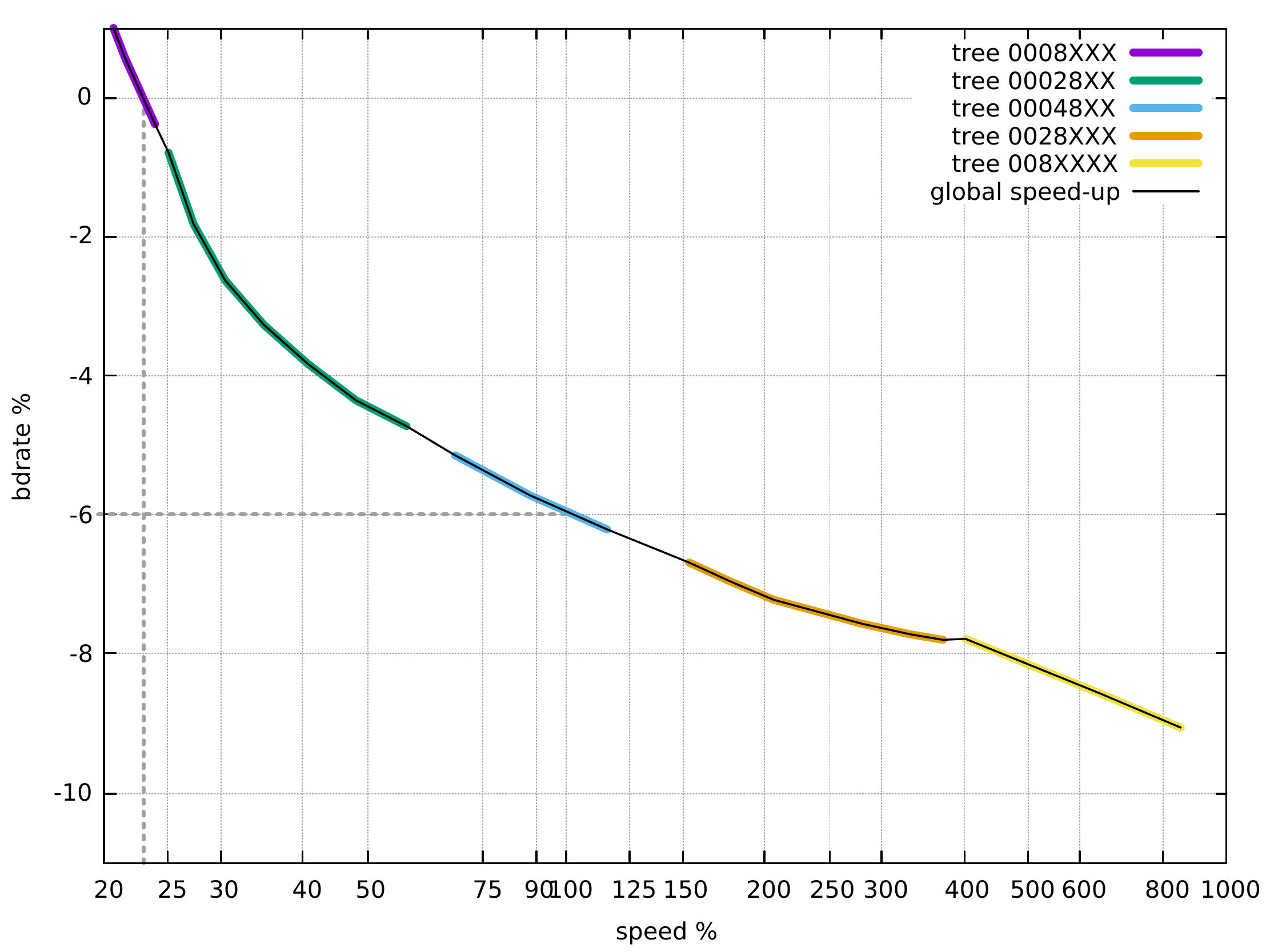}
  \end{minipage}
    \caption{BD-rate gains as a function of the relative encoding time (in \% of JEM7). Left: depending on the tree structure, Right: with dynamic tree structure.}\label{fig:dynamic_tree}
\end{figure}

The left graph of Fig.~\ref{fig:dynamic_tree} shows the trade-off between speed ratio (in $\%$) and BD-rate gain, over JEM7, for different tree structures, as explained below. One can notice the behavior of the encoder when the tree structure is kept and does not depend on the speed-control factor. For low complexity tuning, the QTBT mode is slightly faster at similar BD-rate performance. At similar compression efficiency, the encoding runtime of the codec of \cite{J0022}, configured with QTBT only, is approximately $21\%$ of the JEM7 runtime, while the QTBT+ABT is around $23\%$. On the other hand, QTBT+ABT offers a better trade-off than the QTBT when the speed ratio is above $50\%$ of the JEM7 speed. It is also observed in Fig.~\ref{fig:dynamic_tree} that the maximum gain over JEM is around $4.8\%$ with QTBT only configuration, while it can increase to a maximum gain of $\sim7.3\%$ when extending the partitioning to the QTBT+ABT configuration.
 
On the right graph of Fig.~\ref{fig:dynamic_tree}, the speed-control parameter varies from 0.65 to 3.4 and the tree structure is adapted per range of speed-control values, as can be seen in the varying colors of the plotted line, each color corresponding to a specific tree structure. 
For example, for the legend \textit{tree 028} (in green), corresponding to a speed-up range between $25\%$ and $60\%$:

- levels 0,1 and 2 (QT size 256 to 64) do not allow BT splits (maximum level 0), this does not appear in the legend.

- level 3 (QT size 32) allows a BT tree with a depth of 2; this is indicated by ``2''.

- level 4 (QT size 16) allows a BT tree with a maximum depth of 8, marked ``8''.

- level 5 and 6 (QT size 8 and 4) are not explored by QT.
The other tree types consist of other configurations that can be derived using this nomenclature and the above example. 

It is observed in Fig.~\ref{fig:dynamic_tree} that the proposed split prediction and dynamic tree structure allows the encoder to cover a wide range of trade-offs. For the same BD-rate as JEM7, the encoder runtime is $23\%$ of the JEM runtime, which corresponds to a speed-up of factor $4.3$. At the same speed as JEM7, the BD-rate gain is around $6\%$, and for a speed of approximately $850\%$ of JEM7, the BD-rate gain reaches $9.1\%$.

\section{CNN implementation and discussion on the results}\label{sec:implementation} 

As this algorithm was developed in the context of JVET, some choices were done because of the particular constraints of JVET. First, the CNN was trained using a data set produced using an encoder optimized for the PSNR, as opposed to a perceptually optimized data set. Second, the trade-off between quality and speed is evaluated using a mono-thread, CPU only, stand-alone software, as opposed to a multi-threads or GPU/FPGA based implementation. It means that the trade-off is evaluated by comparing the speed of running a CNN to decide the splits of a block, compared to the speed of running the classical RDO, in the same conditions.
Hence, the size of the chosen CNN model is quite small to run on CPU in a single thread. 

In this context, a stand-alone C\texttt{++} version of the above CNN software has been developed to obtain the presented computer timing results.  The code is available on the JVET web-site as part of the response to the JVET CfP \cite{J0022}. It can be used as an inference engine by the encoder, without any dependency to external library.


As a comparison, for a single HD frame, the implemented GPU version requires around 180ms/frame (on a GTX 1070), the optimized CPU version of Tensorflow on one core around 2s/frame, and our module around 3.4s/frame.
In the encoder, for each new block $64\times 64$ of an intra slice, the model is inferred, taking as inputs: the block $64\times 64$, its causal border and the normalized QP. The resulting boundary probability vector is saved for processing the sub-blocks inside the root-block.

\section{Conclusion}

A deep-learning-based method for driving the partitioning of CTUs, supporting sophisticated structures, has been described. Operating independently on luma and chroma $64\times 64$ root blocks in intra slice, it determines a subset of most probable split decisions to be explored by the classical RDO, speeding-up the encoder. A single parameter drives the proper thresholds and tree structure to find the targeted trade-off between encoder complexity (speed-up) and efficiency (BD-rate gain). Tested on the encoder of the CfP response presented in \cite{J0022}, for a given allowed topology of splits, a speed-up of $\times 2$ is obtained without BD-rate loss, or a speed-up above $\times 4$ with a loss below 1\% in BD-rate, in All Intra configuration. As advanced structures enable solid compression gains, while impacting the complexity at the encoder side only, deep learning offers new perspectives for developing video codecs supporting more elaborated partitioning structures.

\Section{References}
\bibliographystyle{IEEEbib}
\bibliography{main}

\end{document}